# Single Shot Spiral TSE with Annulated Segmentation

Juergen Hennig

Antonia Barhoorn

Shuoyue Zhang

Maxim Zaitsev

Adress for all authors:

University Medical Center FREIBURG, Dept.of Radiology, Medical Physics, Kilianstr.5A, 79106 Freiburg, Germany, www.mr.uniklinik-freiburg.de

Corresponding author:

Juergen Hennig

e-mail:

juergen.hennig@uniklinik-freiburg.de

Abstract

Purpose: To develop, optimize and implement a single shot spiral TSE-sequence at 3 T and to demonstrate its feasibility to acquire artefact free images of the central nervous system with 1 mm spatial resolution in < 200 ms.

**Theory and Methods:** Spiral TSE sequences with annulated spiral segmentation have been implemented with different acquisition modes. In fixed mode the duration of each spiral segment is fixed to fill the available acquisition time $t_{acq}$. In tangential mode, the beginning of each spiral segment is reached via a straight tangential trajectory. Tangential mode allows faster transition and thus longer $t_{acq}$ for a given echo spacing ESP, but less data points can be acquired per acquisition interval. Alternating between spiral-in and spiral-out readout in alternating echoes leads to a somewhat different point spread function for off-resonant spins.

**Results:** Images of the brain with 1 mm spatial resolution acquired with a variable density spiral with ~45 % undersampling can be acquired in a single shot. All acquisition modes produce comparable image quality. Only mild artifacts in regions of strong susceptibility effects can be observed for ESP of 10 ms and below. The use of variable flip angle schemes allows seamless acquisition of consecutive slices and/or dynamic scans without waiting time between consecutive acquisitions. Comparison with images acquired at 1.5 T shows reduced susceptibility artifacts within the brain and facial structures..

**Conclusion:** Single shot spiral TSE has been demonstrated to enable highly efficient acquisition of high-resolution images of the brain in < 200 ms per slice.

Introduction

The possibility to use a spiral k-space trajectory to cover k-space for MR imaging has been described as early as 1983 by Ljunggren (1). First experimental demonstration of MR images followed in 1986 (2). In the 90s intense development took place, mainly driven by the Stanford group around G. Glover, G. Meyer, and A. Macovski (3–8). Spiral trajectories offer a number of attractive features. Due to the isotropic point spread function (PSF), no spatial misregistration occurs and the trajectory is periodically self-compensating for flow and motion. Spirals also offer the most efficient way to cover k-space for current gradient system with performance limited by slew rate and gradient amplitude. In spite of its advantages spiral imaging has not made it into clinical routine so far. This is due to the challenges associated with spiral scanning: Deviations of the executed k-space trajectory from the nominal trajectory lead to image artifacts. Commonly encountered imperfections include non-zero delay times of the gradients, imperfections of the gradient hardware, eddy current effects, field inhomogeneities, and concomitant fields. Even if gradient waveforms are executed perfectly, off-resonance effects lead to a continuous dephasing along the spiral trajectory, which produce phase discontinuities across k-space with resulting artifacts. Additionally, local susceptibility-induced gradient fields lead to distortions of the effective encoding trajectory (9).

Hardware dependent imperfections like concomitant fields, gradient delays and – to some extent – eddy current effects remain constant over time and can be corrected by proper calibration (10–17). Most elegantly this is performed by measuring the true encoding trajectory with a field camera, which has shown to allow spiral imaging at 7 T (18).

Subject dependent effects are more tricky to handle and may require additional calibrations like a field map for off-resonance and inhomogeneity correction (17,19). The sensitivity to off-resonance effects requires fat suppression in most parts of the body. Correction methods of varying degree of sophistication have been proposed for all relevant imperfections, but these mostly require additional effort and always carry the risk of residual artifacts. A minor misadjustment of the resonance frequency, which is barely noticeable in standard Fourier imaging techniques, may lead to drastic artifacts in spiral imaging. Field inhomogeneities lead to a benign signal shift in FT imaging, but no loss of e.g. sharp structures, whereas the isotropic PSF of spirals transforms the frequency offset to blurring and loss of signal - to name just some of the main challenges.

The use of spiral imaging in combination with a multi-spin echo sequence (RARE, FSE, TSE,… - in the following TSE is used) has been introduced in 1997 by W. Block (20). The combination of a spiral trajectory with a multiecho readout necessitates a segmented spiral approach. In order to maintain a smooth amplitude modulation along the trajectory and thus across k-space, an annular segmentation scheme was proposed in which the central circle of k-space is read out in the first echo and continuous ring-shaped segments are consecutively acquired in subsequent echoes.

More recently intense new developments in spiral TSE have been performed by various groups (8,23–27). For 2D spiral TSE (23) a standard interleaved approach was suggested by Zhiqiang Li and Jim Pipe, in which one interleaf is read out in each echo. In order to read the center of k-space out - which is acquired in each echo - at the center of each refocusing period, each interleaf was realized as an in-out spiral trajectory. Reordering the interleaves leads to a strong tangential sawtooth modulation of k-space signals. The resulting artifacts are minimized by double encoding – i.e. acquisition with opposite direction of each interleaf – and demodulation with a filter function, where the filter is optimized for a mean targeted T2 of the tissue. The clinical usefulness of the sequence was recently demonstrated (26).

The purpose of the present study was to implement and optimize a single shot spiral TSE sequence for rapid imaging. The target was to realize proton density (PD) and T2-weighted imaging with 1 mm isotropic resolution without the necessity for correction by additional means like field maps. Implementations were performed at 3 and 1.5 T.

Methods

2.1. Segmented Spiral Trajectories

Spiral trajectories were created with the variable density spiral toolbox from Brian Hargreaves (27,28). Typically, variable density spirals were used such that the total length of the trajectory was reduced to ~60% compared to a fully resolved single density spiral.

In order to maintain the CPMG-condition the k-space trajectory has to be brought to the same point $K_{ref+}$ in k-space before each refocusing pulse, which for conventional rectilinear implementation is achieved by the phase encoding rewinder pulse (29). The refocusing pulse will the mirror $K_{ref+}$ to its mirror point $K_{ref-}$. In the original paper, Block has used the k-space origin as refocusing point $K_{ref+/-}$ (20). This may carry the risk that for refocusing flip angles deviating from 180° a free induction decay will be created by each refocusing pulse, which follows a nominally identical k-space trajectory as the proper signal, but which will be affected differently by off-resonance effects. The FIDs will be refocused at the beginning, whereas for echoes refocusing will occur at the center of each readout. As a result, image artifacts may be created by interference of the two signals. Therefore – just like in standard RARE - a prefocusing gradient is applied between the excitation pulse and the first refocusing pulse, which puts $K_{ref}$ outside of the k-space center and preferably outside of the part of the k-space sampled by the spiral trajectory.

Fig. 1 shows the basic pulse diagram for the sequence. The time tsp before and after each refocusing pulse is used to apply a spoiler gradient in the slice selection direction, it also gives the time required to reach the beginning of each spiral segment from $K_{ref-}$ .

There are two basic ways to realize this transition:
- In a fixed mode the number of points in each spiral segment is kept constant throughout the echo train. This means that the starting- and end-points of the various segments can be anywhere in the circular part of k-space covered by the spiral trajectory. The time $t_{sp}$ allocated to cover this transition therefore needs to be sufficiently long to cover this transition trajectory for the given maximum amplitude and slew rate of the available gradient system.
- In tangential mode the initial trajectory meets the next spiral segment tangentially and also leaves it tangentially after the end of each spiral segment. In this mode the number of data points in each spiral segment will vary and some parts of the spiral trajectory will be scanned twice.

Fig. 2 shows example trajectories for both above-mentioned modes. It is clear, that the tangential mode (Fig. 2b) allows for much shorter $t_{sp}$ compared to the fixed mode (Fig. 2a). The minimum $t_{sp}$ that can be realized depends on the time necessary to reach the respective starting point of each segment. This transition trajectory should be such that it smoothly merges with the following spiral segment, i.e. gradient vector components at the meeting point should be identical to the instantaneous gradient components at the start of the following spiral segment. It is clear that sampling in tangential mode is less efficient compared to the fixed mode since each spiral segment has to be entered and left at the specific points compatible with a tangential transition. Therefore, the number of points in each segment varies and has to be less than the maximum number of data points possible within the acquisition time $t_{acq}$.

Fig. 3 shows a color-coded map of the minimum time necessary for the transition trajectory to reach individual points on a spiral designed for a maximum slew rate of 150 T/m/s and a maximum amplitude of 35 mT/m. The transition trajectories (not shown) are calculated such that the respective point on the spiral is reached with the same speed and amplitude as the ensuing spiral as suggested in Fig. 2. The k-space coordinates are given in spatial units (1/m), the k-space width (-500….500) corresponds to 1 mm spatial resolution. The trajectory has been regridded to allow a smooth color representation. The map illustrates that for a counterclockwise trajectory k-space points in the bottom half can be reached faster compared to their positive mirror points, for which the transition trajectory has to partly reverse direction in order to achieve a smooth transition. It is demonstrated that the maximum time necessary to reach the most inaccessible point on the spiral is ~1.05 ms, which is then chosen as $t_{sp}$ for all echoes.

In tangential acquisition mode the maximum time necessary to get to each spiral segment is much shorter. As shown in Fig. 2 the transition points lie at the intersection of the spiral with a semicircle over the line from $K_{ref}$ to the center of the spiral indicated by the white semicircle in Fig. 3. The maximum transition time is shorter than $t_{sp}$=0.6 ms. The shorter $t_{sp}$ compared to the fixed mode allows to somewhat compensate for the less efficient sampling in tangential mode. Supplementary Tab. 1 shows the comparison of the efficiency of sampling between both modes for the gradient specifications (Gmax= 35 mT/m, slew rate= 150 T/m/s) used in our experiments at 3 T. Simulations have been made for a field-of-view of 24 cm, 1 mm spatial resolution and a variable density spiral leading to a mild reduction of the total number of sampled points to 56 % compared to a fully sampled spiral. The absolute numbers regarding ETL and ETD will of course vary with the exact parameters used.

The respective map for the original sequence (20) with $K_{ref}$ at the center of k-space is also shown for comparison. The fourfold symmetry of the map is a consequence of the fact that the maximum slew rate is set for each gradient individually i.e. regions in k-space where both gradients work optimally together can be reached faster. The maximum transition time is only slightly longer than that for the tangential mode.

For on-resonant spins $T_2$-decay leads to an isotropic attenuation of the signal falling away from the k-space center with some minor steps between successive segments due to T2-decay during the time between successive segments. $T_2^*$-effects would lead to additional signal modulation but are negligible for the echo spacings used in our experiments. This leads to an isotropic T2-dependent blurring. Off-resonance effects are refocused at the center of each echo and lead to symmetric dephasing towards the beginning and end of each acquisition time. This leads to a periodic signal modulation along the echo train (Fig. 4) which translates into a ring-like modulation in k-space (s. Fig. S1 in supplementary materials). The resulting sidebands in the PSF increase with off-resonance frequency (see also discussion in (20)). The distance of the main sideband peak from the center is given by the number of echoes. For tacq= 5 ms, $2\pi$ dephasing corresponds to an off-resonance frequency of 200 Hz. A field map of the head of a normal volunteer (Fig. 4b) shows that the maximum susceptibility dependent off-resonance frequency can be as high as +-300 Hz, but the extrema lie outside of the CNS except for the frontal regions above the sinuses indicated by red arrows.

As a more subtle point it is noted that the off-resonance dependent dephasing at the beginning and end of each segment has opposite phase which leads to a phase jump when joining the data for reconstruction. This can be avoided if data acquisition is performed in an alternating mode in which signal readout alternates from spiral-in to spiral-out in alternating echoes. It can be shown that this leads to a slightly different PSF with identical intensity of the central peak but different distribution of sideband intensities (s. Fig. S2 in supplementary materials).

2.2. Optimization of data acquisition

In the original paper (20) longer effective echo times $TE_{eff}$ needed to generate T2-contrast have been achieved by starting the readout of the spiral at a later time point matching the echo time, thus leaving the previous refocusing periods unused. We have used a different approach, similar to the standard Cartesian TSE, in which longer $TE_{eff}$ is achieved by a cyclic permutation of the individual spiral segments.

The echo train duration ETD suggests the use of lower refocusing flip angles in order to slow down signal decay by simulated echo contributions (30–33). This has been achieved using a TRAPS approach (32) with a smooth decay of flip angles over the first 4-5 echoes (s. Fig. S3 in supplementary materials). Signal intensities with predefined T1 and T2 have been calculated using the Extended Phase Graph algorithm (34,35) and used to normalize the measured signals . As a side effect low refocusing flip angles lead to a reduction of the specific absorption rate SAR by a factor of 3-8, which is especially beneficial for multislice and/or dynamic acquisitions at 3 T.

2.3. Implementation

Single shot Spiral TSE has been implemented at 3 T (Magnetom PrismaFit, Siemens Healthineers, Erlangen) as well as 1.5 T (Magnetom Aera, Siemens Healthineers, Erlangen). Gradient specifications used to design the pulse sequences were deliberately reduced with respect to the maximum hardware specifications to avoid peripheral nerve stimulation (PNS) and were at 3T were Gmax=35 mT/m and slew rate = 150 T/m/s and Gmax=30 mT/m and slew rate = 125 T/m/s at 1.5 T, respectively. Sequences were implemented using the Pulseq sequence programming environment (36,37), which is distributed via GitHub (https://github.com/pulseq/pulseq). Image reconstruction was performed using the non-uniform FFT (*nufft*) of the BART Toolbox (38). Iterative reconstruction with the *pics* algorithm gives slightly better results, but takes much longer; therefore it has been used only selectively. Most image reconstructions have been performed offline, however, recently BART reconstruction has been implemented on the 3 T scanner using the 'Framework for Image Reconstruction Environments (FIRE) for Advanced Reconstruction Prototyping' provided by the manufacturer (39).

All experiments were performed in accordance with the protocol approved by the local IRB of our institution. Experiments have been performed on 18 healthy volunteers (17 male, 1 female) at 3 T, 3 Volunteers were also scanned at 1.5 T. Written informed consents were acquired from all volunteers prior to the experiments.

3. Results

Fig. 5 shows 8 out of 15 slices acquired in a multislice single shot spiral TSE-experiment at 3T (echo spacing ESP 8 ms, $t_{acq}$ = 3.9 ms, $TE_{eff}$ 72 ms, slice thickness SLTH 5 mm, slice gap 8 mm). Image reconstruction has been performed using the nonlinear reconstruction of BART (details in supplementary materials) after renormalization of the echo amplitudes using signal amplitudes calculated with the extended phase graph algorithm for $T_1$ = 1s, $T_2$ = 200 ms (s. Fig. S4 in supplementary materials). Good image quality with good gray matter/white matter contrast is observed throughout the brain including the cerebellum. Some minor shading can be perceived in the frontal brain above the sinuses (s. arrow), otherwise the CNS (including the entire brain, the cerebellum) is depicted clearly and sharp. Facial structures visible in the lower slices appear blurred.

Fig. 6 shows images with different T2 contrast and demonstrate that the permutation of the spiral segments to achieve higher $TE_{eff}$ does not induce image artifacts.

Fig. 7 demonstrates images acquired at identical positions with 4 different acquisition modes: fixed (A,B) tangential (C,D) mode, respectively, with a continuous spiral readout (A,C) and alternating spiral in-out and out-in in subsequent echoes (B,D), respectively. Normalization has been adjusted to achieve identical k-space filtering of acquired data. Image contrast is slightly different due to the different echo train length of tangential vs. fixed acquisition modes (s. Supplementary Tab. 1). It is shown that all acquisition modes produce artefact-free images of comparable quality. Tangential modes require longer echo trains (ETL=32 compared to 20 for fixed mode) but are consistently shown to be somewhat sharper compared to fixed acquisition modes (this is also true for non-normalized images, data not shown). This may be due to slight deviations of the actual trajectory compared to its nominal value in fixed mode acquisition caused by the much more demanding initialization trajectory compared to the smooth transition in tangential mode as shown in Fig. 2.

The effect of spurious echoes caused by insufficient spoiling of spurious FIDs is shown in Fig. 8. An image acquired with $K_{ref}$ at kx, ky = -550,0 1/m (A) is taken as reference, (B)-(D) show images acquired with $K_{ref}$ = 0,0 (B), -50,0 (C), and -100,0 (D) 1/m respectively. Images in the bottom row show the difference to the reference image (A). The interference stripes on the difference images clearly demonstrate the interference of the proper signals with signals arising from the FIDs generated by the refocusing pulses. Interference of the desired and spurious signals is not obvious in image (B), however, upon a careful comparison with image (A) image intensity and contrast alterations can be discovered.

Fig. 9 shows a comparison of images acquired with different echo spacing. It is demonstrated that blurring in the frontal brain above the sinuses increases with echo spacing ESP (and thus acquisition time $t_{adc}$). At longer ESP overall image quality deteriorates quickly as expected from the field map (Fig. 5)

Finally, Fig. 10 shows a comparison between of selected slices acquired on the same volunteer at 3 T vs. 1.5 T with identical acquisition parameters (ESP 10 ms, tangential mode, $TE_{eff}$ 70 ms). Images demonstrate reduced susceptibility dependent blurring of the 1.5 T images (yellow arrows). Facial structures (red arrows) still appear blurred at 1.5 T but less so compared to 3 T. The 1,5 T images are visibly more noisy. The noise level and characteristics very much depend on the actual parameters used in the BART-reconstruction, so we did not perform any quantitative SNR comparison so far.

Discussion

Our results show, that single shot spiral TSE is feasible at 3 T. Good image quality is demonstrated for tacq <= 6 ms, which corresponds to ESP<= 10 ms in our sequences. Fat suppression is essential in order to avoid image artifacts. Data acquisition for images with 1 mm in plane resolution can be realized with an echo train duration of around 200 ms using a variable density spiral with undersampling of ~ 56 %. Even shorter echo trains may be realized using more aggressive undersampling, a detailed discussion of different undersampling strategies is beyond the scope of this paper.

The use of variable flip angle schemes according to the principles of TRAPS (32) leads to a flattening and prolongation of the signal decay curve. For a targeted flip angle of 60° acquisition can be performed continuously without SAR problems even at 3 T, without any waiting time between the acquisition of successive slices and/or successive scans. Whole brain coverage with 20-30 slices can thus be achieved in ~3-8 s including fat suppression before each slice.

Normalization of raw data based on simulated echo envelopes with predefined $T_1$ and $T_2$ improves image quality and avoids blurring.

Setting the refocusing point $K_{ref+/-}$ outside of the acquired k-space is a prerequisite for the use of the tangential acquisition mode. It also avoids interference of the proper signal with spurious echoes. For fixed mode acquisition the amplitude of the resulting interference could in principle also be reduced by increasing the spoiler gradients placed symmetrically around the refocusing pulses.

Tangential acquisition mode is furthermore advantageous in term of the probability of inducing PNS or avoiding PNS limits. It is well-known that the PNS limitations the most stringent for the Y physical gradient axis (assuming the patient is in the supine position)(40). This means that the same gradient pulses, that can be safely played out on X or Z axes can lead to subject stimulation if applied to the Y axis. For the tangential acquisition mode it is possible to orient the logical acquisition coordinates such that the transition segments of the trajectory can be predominantly oriented along the directions less limited by the PNS constraints. This is not the case for the fixed acquisition mode, where the transition segments can occur in any direction within the imaging slice.

Images with different effective echo times can be realized by cyclic permutation of the individual spiral segments along the echo train, similar to a linear phase encoding scheme in Cartesian RARE(TSE, FSE,…)(41).

The single shot spiral TSE technique by Zhiqiang Li (23,26) uses a interleaved spiral segmentation with double encoding. Reconstruction is performed with off-resonance correction using a field map leading to excellent image quality in ~ 2 min acquisition time. The focus of our work was on single shot acquisition without any additional signal processing, so the two techniques can be considered complimentary and work in different application domains. Off-resonance correction would certainly also improve image quality of our images, but there doesn't seem to be a point to spend several minutes to acquire a field map just to save a few seconds for the actual acquisition.

The corresponding Cartesian sequence is single shot RARE, a.k.a. HASTE (42). A single shot implementation with 1 mm resolution at a similar acceleration factor would require rather long echo trains, which necessitate low refocusing flip angles leading to reduced SNR and loss of gray matter/white matter contrast. On the other hand, Cartesian k-space coverage is very robust against field inhomogeneities and frequency misadjustments. Applications outside of the CNS are straightforward to realize with HASTE, but unfeasible with single shot spiral TSE. Use cases for the proposed technique may be applications, were the very short overall acquisition times are essential, e.g. for non-cooperative patients as an alternative to PROPELLER TSE (43) or for motion monitoring by continuous MR acquisition as used in hybrid PET-MR imaging (44).

More widespread applications will become feasible at lower fields like the recently introduced 0.55 T systems (45) provided that they come with high-performance gradients. In the regime of the spiral trajectories used, performance is limited by slew rate rather than maximum amplitude. Reducing the slew rate from 150 T/m/s to 50 T/m/s leads to prolongation of the echo train at ESP=10 ms from 130 ms to 310 ms, which is not compatible with single shot acquisition anymore. For implementation of the sequence at low fields using high-performance gradients concomitant gradients may have to be considered (12) and corrected for according to the principles shown in (46).

Data Availability Statement

The source code to produce the single shot spiral TSE sequences for use with the Pulseq-sequence programming environment is available in https://github.com/HennigJue/single-shot-spiral-TSE.

References

1. Ljunggren S. A Simple Graphical Representation of Fourier-Based Imaging Methods. Journal of Magnetic Resonance 1983;54:338–343 doi: 10.1016/0022-2364(83)90060-4.

2. Ahn CB, Kim JH, Cho ZH. High-speed spiral-scan echo planar NMR imaging-I. IEEE Trans Med Imaging 1986;5:2–7 doi: 10.1109/TMI.1986.4307732.

3. Glover GH. Simple analytic spiral K-space algorithm. Magn Reson Med 1999;42:412–415 doi: 10.1002/(sici)1522-2594(199908)42:2<412::aid-mrm25>3.0.co;2-u.

4. Yen YF, Han KF, Daniel BL, et al. Dynamic breast MRI with spiral trajectories: 3D versus 2D. JMRI-J. Magn. Reson. Imaging 2000;11:351–359 doi: 10.1002/(SICI)1522-2586(200004)11:4<351::AID-JMRI2>3.0.CO;2-L.

5. Glover GH, Lai S. Self-navigated spiral fMRI: Interleaved versus single-shot. Magn. Reson. Med. 1998;39:361–368 doi: 10.1002/mrm.1910390305.

6. Daniel BL, Yen YF, Glover GH, et al. Breast disease: Dynamic spiral MR imaging. Radiology 1998;209:499–509 doi: 10.1148/radiology.209.2.9807580.

7. Meyer CH, Hu BS, Nishimura DG, Macovski A. Fast Spiral Coronary Artery Imaging. Magnetic Resonance in Medicine 1992;28:202–213 doi: 10.1002/mrm.1910280204.

8. Fielden SW, Mugler JP, Hagspiel KD, Norton PT, Kramer CM, Meyer CH. Noncontrast Peripheral MRA with Spiral Echo Train Imaging. Magn. Reson. Med. 2015;73:1026–1033 doi: 10.1002/mrm.25216.

9. Noll DC. Rapid MR image acquisition in the presence of background gradients. In: Proceedings IEEE International Symposium on Biomedical Imaging. ; 2002. pp. 725–728. doi: 10.1109/ISBI.2002.1029360.

10. Brodsky EK, Klaers JL, Samsonov AA, Kijowski R, Block WF. Rapid measurement and correction of phase errors from B0 eddy currents: Impact on image quality for non-cartesian imaging. Magnetic Resonance in Medicine 2013;69:509–515 doi: 10.1002/mrm.24264.

11. Harkins KD, Does MD, Grissom WA. Iterative Method for Predistortion of MRI Gradient Waveforms. IEEE Trans. Med. Imaging 2014;33:1641–1647 doi: 10.1109/TMI.2014.2320987.

12. King KF, Ganin A, Zhou XHJ, Bernstein MA. Concomitant gradient field effects in spiral scans. Magn. Reson. Med. 1999;41:103–112.

13. Lechner SM, Sipilae PT, Wiesinger F, Kerr AB, Vogel MW. Spiral Imaging Artifact Reduction: A Comparison of Two k-Trajectory Measurement Methods. J. Magn. Reson. Imaging 2009;29:1485–1492 doi: 10.1002/jmri.21782.

14. Middione MJ, Loecher M, Moulin K, Ennis DB. Optimization methods for magnetic resonance imaging gradient waveform design. NMR Biomed.:e4308 doi: 10.1002/nbm.4308.

15. Robison RK, Li Z, Wang D, Ooi MB, Pipe JG. Correction of B0 eddy current effects in spiral MRI. Magn Reson Med 2019;81:2501–2513 doi: 10.1002/mrm.27583.

16. Tan H, Meyer CH. Estimation of k-space trajectories in spiral MRI. Magn Reson Med 2009;61:1396–1404 doi: 10.1002/mrm.21813.

17. Delattre BMA, Heidemann RM, Crowe LA, Vallee J-P, Hyacinthe J-N. Spiral demystified. Magn. Reson. Imaging 2010;28:862–881 doi: 10.1016/j.mri.2010.03.036.

18. Engel M, Kasper L, Barmet C, et al. Single-shot spiral imaging at 7T. Magn. Reson. Med. 2018;80:1836–1846 doi: 10.1002/mrm.27176.

19. Yudilevich E, Stark H. Spiral Sampling in Magnetic Resonance Imaging-The Effect of Inhomogeneities. IEEE Transactions on Medical Imaging 1987;6:337–345 doi: 10.1109/TMI.1987.4307852.

20. Block W, Pauly J, Nishimura D. RARE spiral T-2-weighted imaging. Magn.Reson.Med. 1997;37:582–590 doi: 10.1002/mrm.1910370418.

21. Frank LR, Jung Y, Inati S, Tyszka JM, Wong EC. High efficiency, low distortion 3D diffusion tensor imaging with variable density spiral fast spin echoes (3D DW VDS RARE). Neuroimage 2010;49:1510–1523 doi: 10.1016/j.neuroimage.2009.09.010.

22. Li Z, Wang D, Robison RK, et al. Sliding-Slab Three-Dimensional TSE Imaging With a Spiral-In/Out Readout. Magn. Reson. Med. 2016;75:729–738 doi: 10.1002/mrm.25660.

23. Li Z, Karis JP, Pipe JG. A 2D spiral turbo-spin-echo technique. Magn Reson Med 2018;80:1989–1996 doi: 10.1002/mrm.27171.

24. Li Z, Pipe JG, Ooi MB, Kuwabara M, Karis JP. Improving the image quality of 3D FLAIR with a spiral MRI technique. Magn Reson Med 2020;83:170–177 doi: 10.1002/mrm.27911.

25. Li Z, Schär M, Wang D, et al. Arterial spin labeled perfusion imaging using three-dimensional turbo spin echo with a distributed spiral-in/out trajectory. Magnetic Resonance in Medicine 2016;75:266–273 doi: 10.1002/mrm.25645.

26. Sartoretti E, Sartoretti-Schefer S, Smoorenburg L van, et al. Spiral 2D T2-Weighted TSE Brain MR Imaging: Initial Clinical Experience. American Journal of Neuroradiology 2021 doi: 10.3174/ajnr.A7299.

27. Lee JH, Hargreaves BA, Hu BS, Nishimura DG. Fast 3D imaging using variable-density spiral trajectories with applications to limb perfusion. Magnetic Resonance in Medicine 2003;50:1276–1285 doi: 10.1002/mrm.10644.

28. Hargreaves BA. Variable-Density Spiral Design Functions.

29. Hennig J, Nauerth A, Friedburg H. Rare Imaging - a Fast Imaging Method for Clinical Mr. Magn.Reson.Med. 1986;3:823–833 doi: 10.1002/mrm.1910030602.

30. Hennig J. Multiecho Imaging Sequences with Low Refocusing Flip Angles. Journal of Magnetic Resonance 1988;78:397–407 doi: 10.1016/0022-2364(88)90128-X.

31. Mugler JP, Wald LL, Brookeman JR. T2-Weighted 3D Spin-Echo Train Imaging of the Brain at 3 Tesla: Reduced Power Deposition Using Low Flip-Angle Refocusing RF Pulses. Proceedings of the International Society for Magnetic Resonance in Medicine 2001;9:438.

32. Hennig J, Weigel M, Scheffler K. Multiecho sequences with variable refocusing flip angles: Optimization of signal behavior using smooth transitions between pseudo steady states (TRAPS). Magn. Reson. Med. 2003;49:527–535 doi: 10.1002/mrm.10391.

33. Hennig J, Weigel M, Scheffler K. Calculation of flip angles for echo trains with predefined amplitudes with the extended phase graph (EPG)-algorithm: Principles and applications to hyperecho and TRAPS sequences. Magn. Reson. Med. 2004;51:68–80 doi: 10.1002/mrm.10658.

34. Hennig J. Echoes—how to generate, recognize, use or avoid them in MR-imaging sequences. Part I: Fundamental and not so fundamental properties of spin echoes. Concepts Magn. Reson. 1991;3:125–143 doi: 10.1002/cmr.1820030302.

35. Weigel M. Extended Phase Graphs: Dephasing, RF Pulses, and Echoes - Pure and Simple. J. Magn. Reson. Imaging 2015;41:266–295 doi: 10.1002/jmri.24619.

36. Layton KJ, Kroboth S, Jia F, et al. Pulseq: A rapid and hardware-independent pulse sequence prototyping framework. Magn Reson Med 2017;77:1544–1552 doi: 10.1002/mrm.26235.

37. Ravi KS, Potdar S, Poojar P, et al. Pulseq-Graphical Programming Interface: Open source visual environment for prototyping pulse sequences and integrated magnetic resonance imaging algorithm development. Magn. Reson. Imaging 2018;52:9–15 doi: 10.1016/j.mri.2018.03.008.

38. Uecker M, Ong F, Tamir JI, et al. Berkeley advanced reconstruction toolbox. Proceedings of the International Society for Magnetic Resonance in Medicine 2015;23:2486.

39. Veldmann M, Ehses P, Chow K, Zaitsev M, Stöcker T. Open-Source MR Imaging Workflow. Proceedings of the International Society for Magnetic Resonance in Medicine 2021:729.

40. Faber SC, Hoffmann A, Ruedig C, Reiser M. MRI-induced stimulation of peripheral nerves: dependency of stimulation threshold on patient positioning. Magnetic Resonance Imaging 2003;21:715–724 doi: 10.1016/S0730-725X(03)00106-1.

41. Hennig J, Nauerth A, Friedburg H. Rare Imaging - a Fast Imaging Method for Clinical Mr. Magn.Reson.Med. 1986;3:823–833 doi: 10.1002/mrm.1910030602.

42. Semelka RC, Kelekis NL, Thomasson D, Brown MA, Laub GA. HASTE MR imaging: description of technique and preliminary results in the abdomen. J Magn Reson Imaging 1996;6:698–699 doi: 10.1002/jmri.1880060420.

43. Forbes KP, Pipe JG, Karis JP, Farthing V, Heiserman JE. Brain imaging in the unsedated pediatric patient: Comparison of periodically rotated overlapping parallel lines with enhanced reconstruction and single-shot fast spin-echo sequences. Am. J. Neuroradiol. 2003;24:796–798.

44. Rakvongthai Y, El Fakhri G. Magnetic Resonance-based Motion Correction for Quantitative PET in Simultaneous PET-MR Imaging. PET Clin 2017;12:321–327 doi: 10.1016/j.cpet.2017.02.004.

45. Campbell-Washburn AE, Ramasawmy R, Restivo MC, et al. Opportunities in Interventional and Diagnostic Imaging by Using High-Performance Low-Field-Strength MRI. Radiology 2019;293:384–393 doi: 10.1148/radiol.2019190452.

46. Mugler JP, Campbell-Washburn AE, Ramasawmy R, Pfeuffer J, Meyer CH. Maxwell Compensation for Spiral Turbo-Spin-Echo Imaging. Proceedings of the International Society for Magnetic Resonance in Medicine 2021:3.

Legends to figures

Figure 1. Sequence diagram for the single shot spiral TSE-sequence. RF denotes RF-pulses, Gx, Gy, and Gz the 3 orthogonal gradients. The time $t_{sp}$ represents the time used for spoiler gradients (in Gz) as well as the time necessary to reach the start of the following spiral segment, $t_{acq}$ is the data acquisition time.

Figure 2. Examples of k-space trajectories for fixed mode (left) respectively tangential mode acquisition. $K_{ref+,-}$ represent the k-space points to be reached before and after each refocusing pulse in order to maintain the CPMG-conditions. In fixed mode each spiral segment has an identical number of data points determined by the acquisition time duration $t_{acq}$. The next segment acquired in the next refocusing period starts where the previous segment ends – indicated by blue circles. In tangential mode the transition trajectories are straight lines entering and leaving the spiral segments tangentially.

Figure 3. Color-coded map of the minimum time necessary to reach each point on the spiral trajectory from the k-space location $K_{ref-}$ for the case when $K_{ref-}$ is placed outside of k-space covered by the spiral (A) and at the center of k-space (B). The trajectory has been regridded for better visualization. The white semicircle connecting $K_{ref+,-}$ with the k-space center represents the data points for tangential mode acquisition.

Figure 4. A) Stacked plot of the point spread function PSF for off-resonance dephasing from 0 to $2\pi$ over $t_{acq}$. For $t_{acq}$ = 5 ms dephasing by $2\pi$ corresponds to 200 Hz. B) field map of a normal volunteer.

Figure 5. Eight out of fifteen slices from a multi-slice acquisition with ESP = 8 ms ($t_{acq}$ = 3.9 ms) acquired at 3 T.

Figure 6. Variation of $TE_{eff}$ by cyclic permutation of the attribution of the spiral segments along the echo train. Data were acquired with ESP = 10 ms ($t_{acq}$ = 5.9 ms) and $TE_{eff}$ = 10 ms (images on top) and 70 ms (bottom) respectively. Refocusing flip angles were 153, 132, 126, 123, 121, 120, 120, 120 …degrees to ensure smooth transition into the pseudo steady state.

Figure 7. Single shot spiral TSE images acquired with fixed (A, B) and tangential (C, D) mode with continuous (A, C) and alternating (C, D) readout of the spiral segments.

Figure 8. Spurious echoes interfere with the proper signals if $K_{ref+,-}$ is placed inside the k-space ranging from -500 to 500 1/m covered by the spiral trajectory for data acquisition with 1 mm spatial resolution; A) shows results of a scan with $K_{ref+,-}$ set to kx, ky= -550,0 1/m, B) for 0,0 (B), -50,0 (C), and -100,0 (D) 1/m, respectively. The bottom row represents difference images to A) scaled by a factor of 5 for better visualization.

Figure 9. Images acquired with different echo spacing: A) 8 (3.9) ms, B) 10 (5.9) ms, C) 12 (7.9) ms, D) 14 (9.9) ms, E) 16 (11.9) and F) 20 (15.9) ms; numbers in brackets represent $t_{acq}$. Artifacts in the frontal brain above the sinuses indicated by the yellow arrow increase with increasing $t_{acq}$.

Figure 10. Comparison of spiral TSE images acquired at 1.5 T (top) and 3 T (bottom) on the same volunteer and with identical sequences. It is demonstrated that images at 1.5 T show reduced artifacts due to susceptibility effects in problematic regions (yellow arrows). Facial structures (red arrows) are better visible but are still strongly blurred at 1.5 T.

Supplementary Materials

Figure S1. A) Modulation of k-space for annulated segmentation with 13 echoes as a function of the dephasing over $t_{acq}$. The real part of the modulation function is shown. The trajectory has been regridded to an image for better visualization. The annulated segmentation leads to a ring-like modulation. B) shows the modulation along a horizontal profile through the k-space center.

Figure S2. Comparison of the point spread function PSF for continuous (left) and alternating (right) acquisition mode. The signal intensity at the center is identical in both modes.

Figure S3. A) Signal decay over 15 echoes with ESP=10 ms for tissue with can be normalized with calculated amplitudes for image reconstruction for tissue with $T_1$=1 and $T_2$= 0.1 s. (B) Refocussing flip angles used for smooth transition into the pseudo steady state according to TRAPS. Specific absorption rate SAR is reduced to ~13%).

Figure S4. Improvement of image quality by the EPG-based normalization. The unfiltered image (A) acquired with refocusing flipangles 126°, 83°, 72°, 66°, 63°, 60°, 60°,….looks slightly blurred due to the initial signal decay compared to images reconstructed from data normalized with a simulated signal decay with T1= 1 s and T2= 200 ms (B) and T1=1s and T2=100 ms (C). The latter roughly corresponds to relaxation parameters for white matter at 3 T (37). The median values (B) were typically used in order to avoid artificial sharpening and/or Gibbs ringing of CSF-filled structures.

Supplementary Table 1: Acquisition parameters $t_{acq}$, echo train length ETL and echo train duration ETD (= ETL * $t_{acq}$) for acquisition of a variable density spiral with spoiler times $t_{sp}$ set to the maximum possible value (1.05 ms for fixed mode and 0.6 ms for tangential mode).

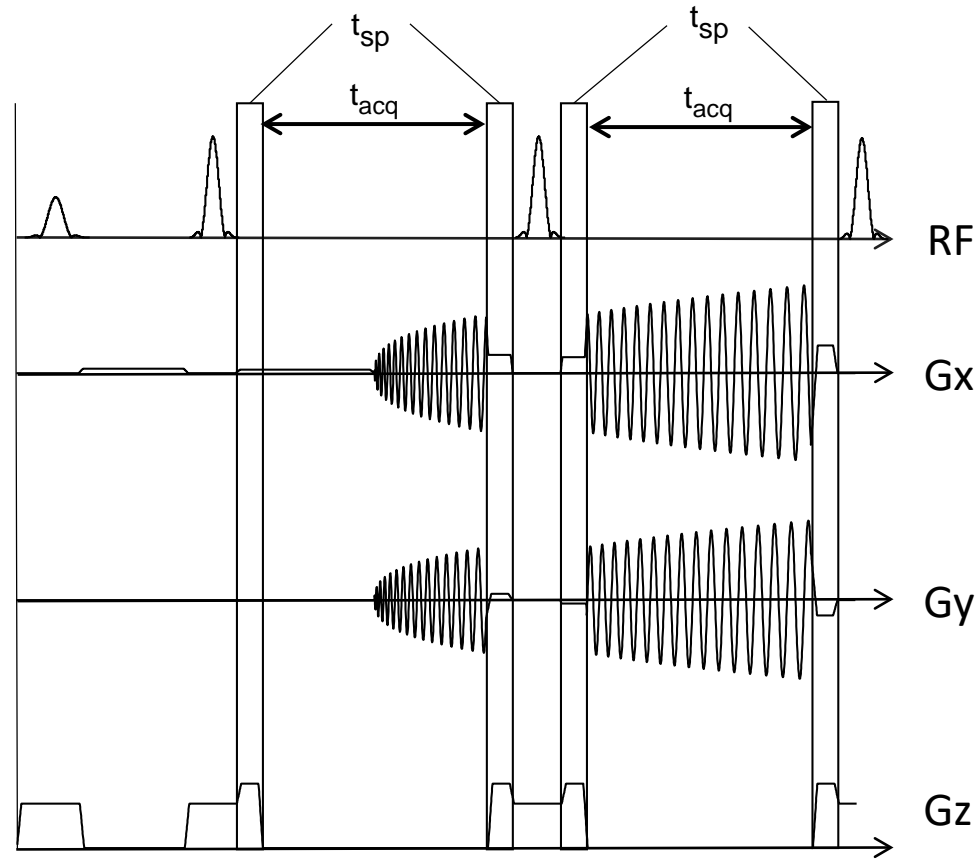

Figure 1. Sequence diagram for the single shot spiral TSE-sequence. RF denotes RF-pulses, Gx, Gy, and Gz the 3 orthogonal gradients. The time $t_{sp}$ represents the time used for spoiler gradients (in Gz) as well as the time necessary to reach the start of the following spiral segment, $t_{acq}$ is the data acquisition time.

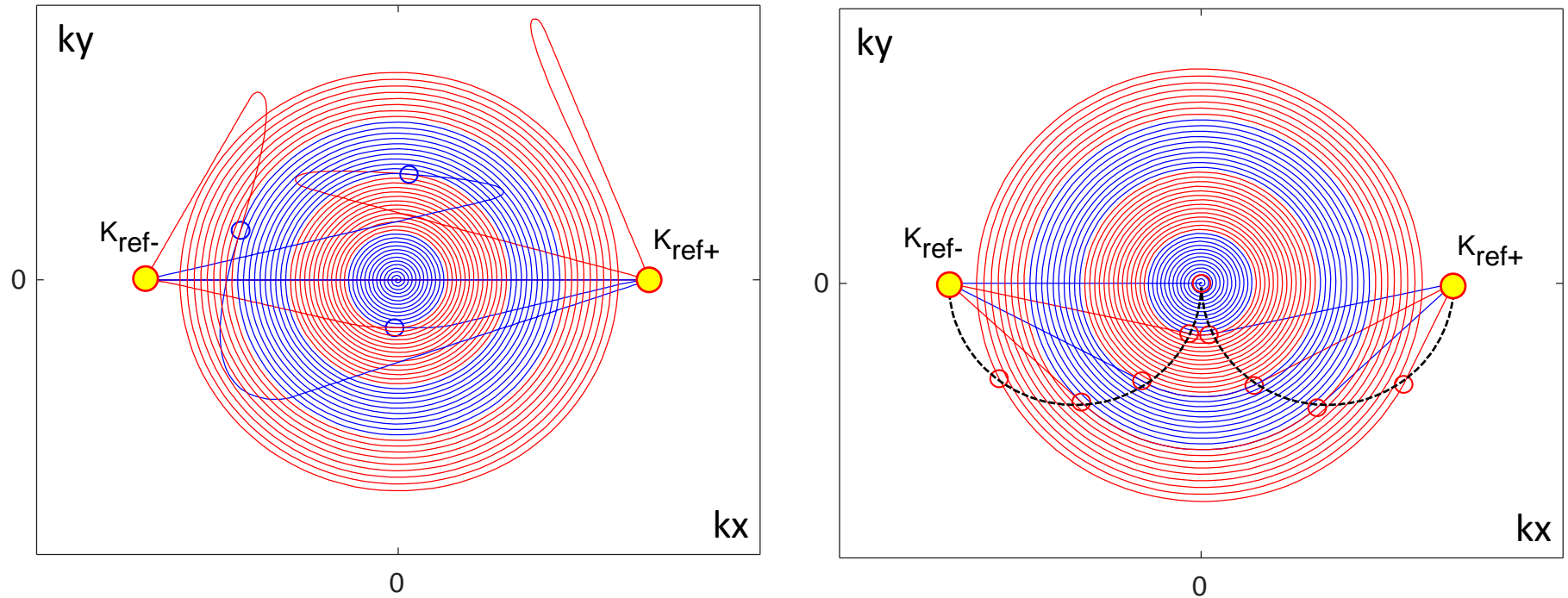

Figure 2. Examples of k-space trajectories for fixed mode (left) respectively tangential mode acquisition. $K_{ref+,-}$ represent the k-space points to be reached before and after each refocusing pulse in order to maintain the CPMG-conditions. In fixed mode each spiral segment has an identical number of data points determined by the acquisition time duration $t_{acq.}$ The next segment acquired in the next refocusing period starts where the previous segment ends – indicated by blue circles. In tangential mode the transition trajectories are straight lines entering and leaving the spiral segments tangentially.

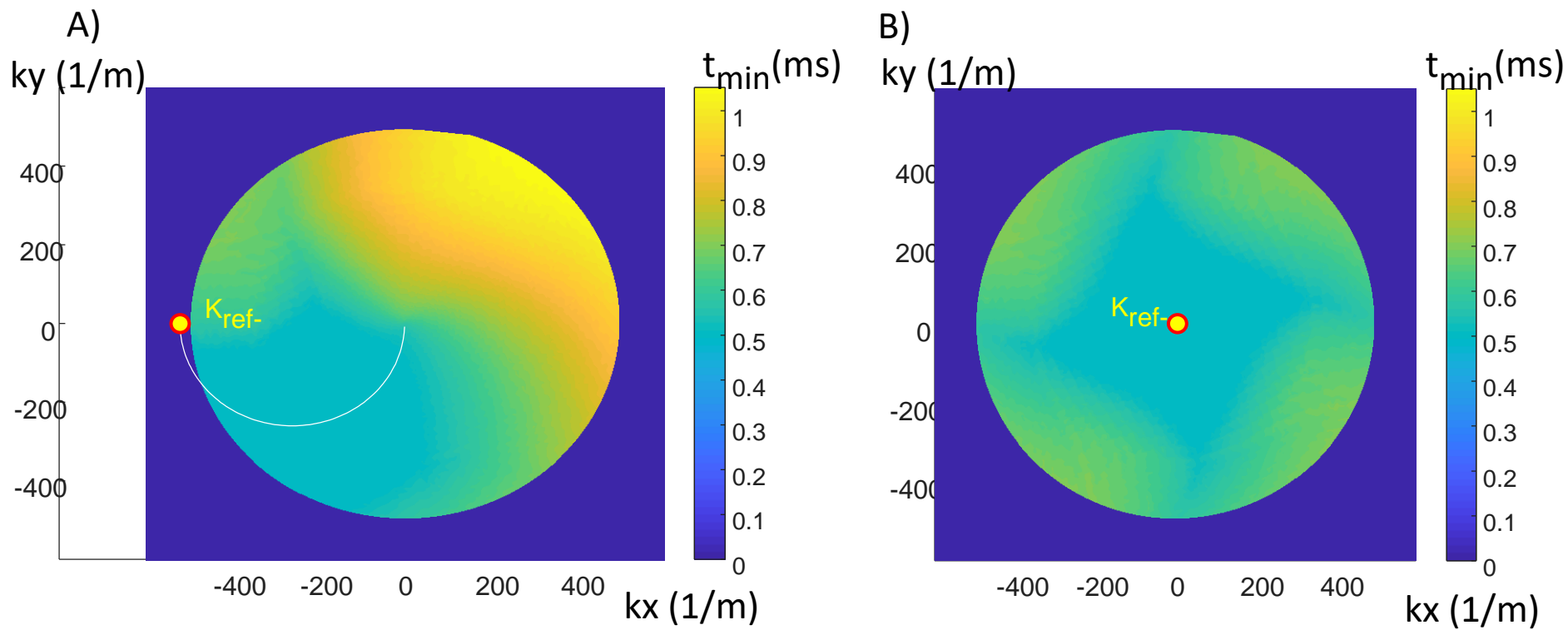

Figure 3. Color-coded map of the minimum time necessary to reach each point on the spiral trajectory from the k-space location $K_{ref-}$ for the case when $K_{ref-}$ is placed outside of k-space covered by the spiral (A) and at the center of k-space (B). The trajectory has been regridded for better visualization. The white semicircle connecting $K_{ref+,-}$ with the k-space center represents the data points for tangential mode acquisition.

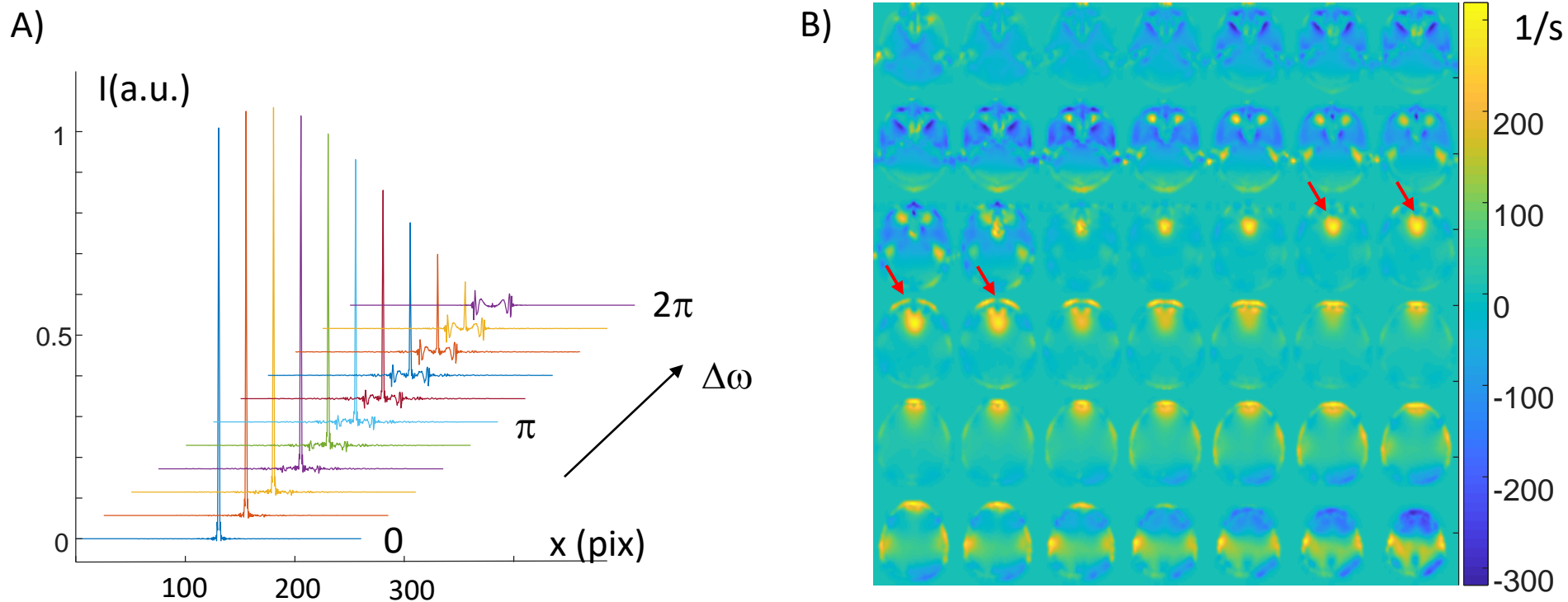

Figure 4. A) Stacked plot of the point spread function PSF for off-resonance dephasing from 0 to $2\pi$ over $t_{acq}$. For $t_{acq}$ = 5 ms dephasing by $2\pi$ corresponds to 200 Hz. B) field map of a normal volunteer.

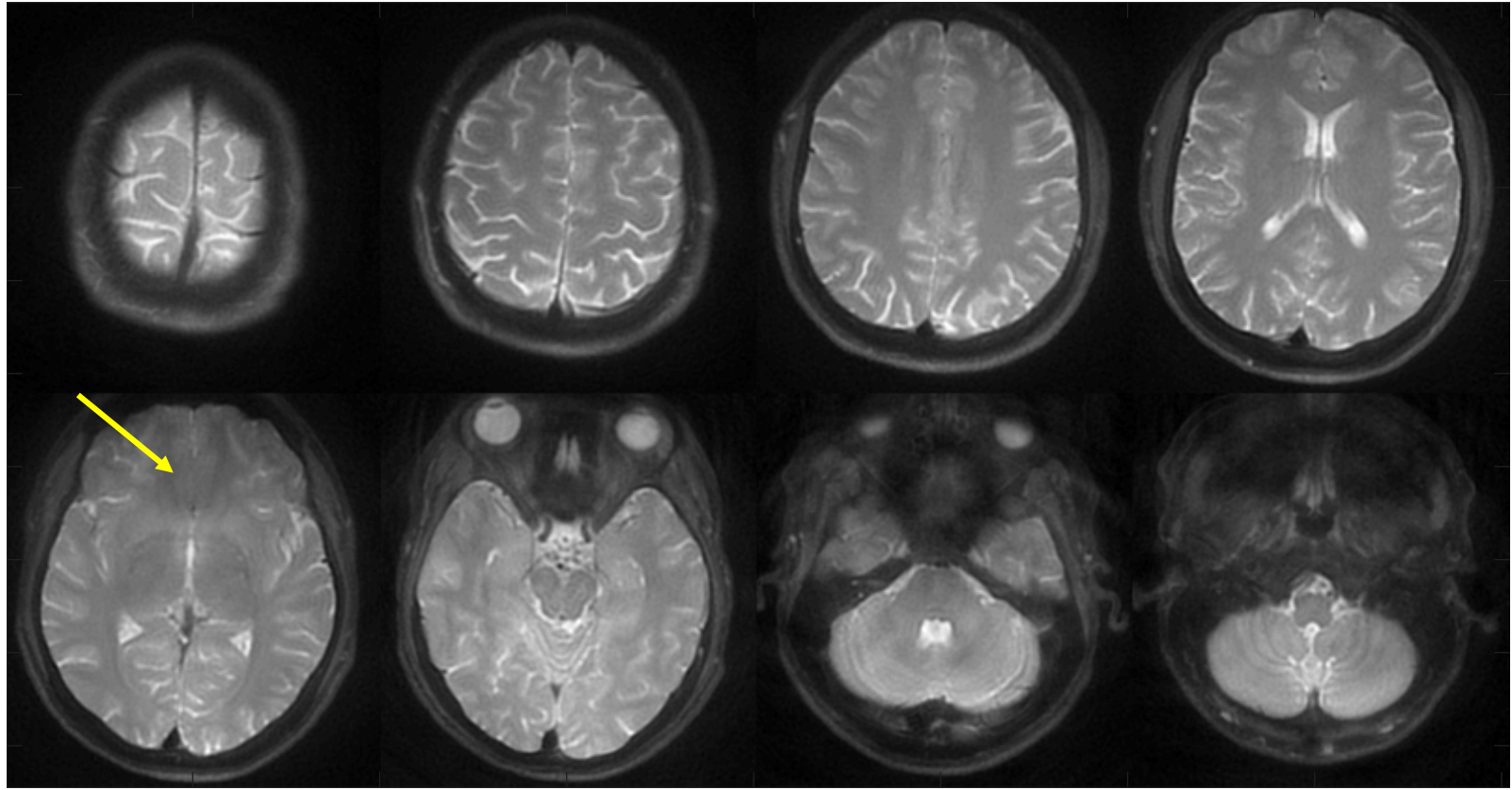

Figure 5. Eight out of fifteen slices from a multi-slice acquisition with ESP = 8 ms ($t_{acq}$ = 3.9 ms) acquired at 3 T.

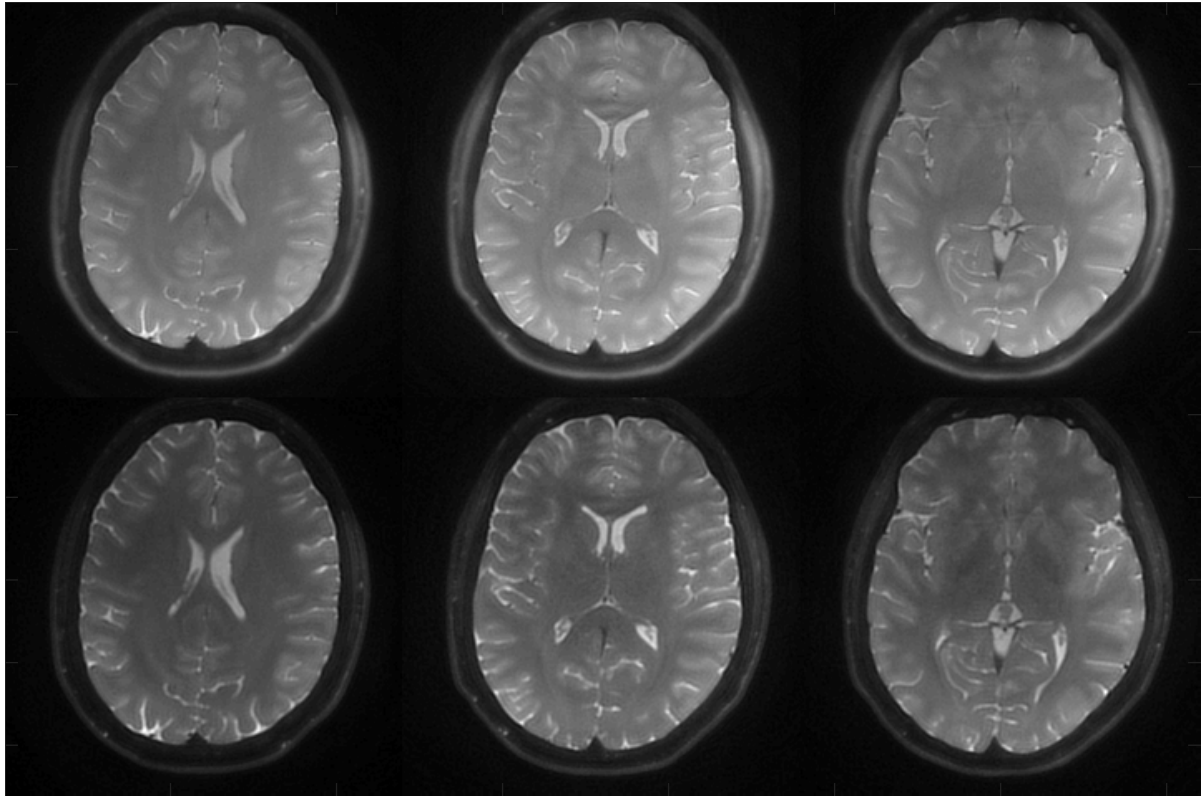

Figure 6. Variation of $TE_{eff}$ by cyclic permutation of the attribution of the spiral segments along the echo train. Data were acquired with ESP = 10 ms ($t_{acq}$ = 5.9 ms) and $TE_{eff}$ = 10 ms (images on top) and 70 ms (bottom) respectively. Refocusing flip angles were 153, 132, 126, 123, 121, 120, 120, 120 …degrees to ensure smooth transition into the pseudo steady state.

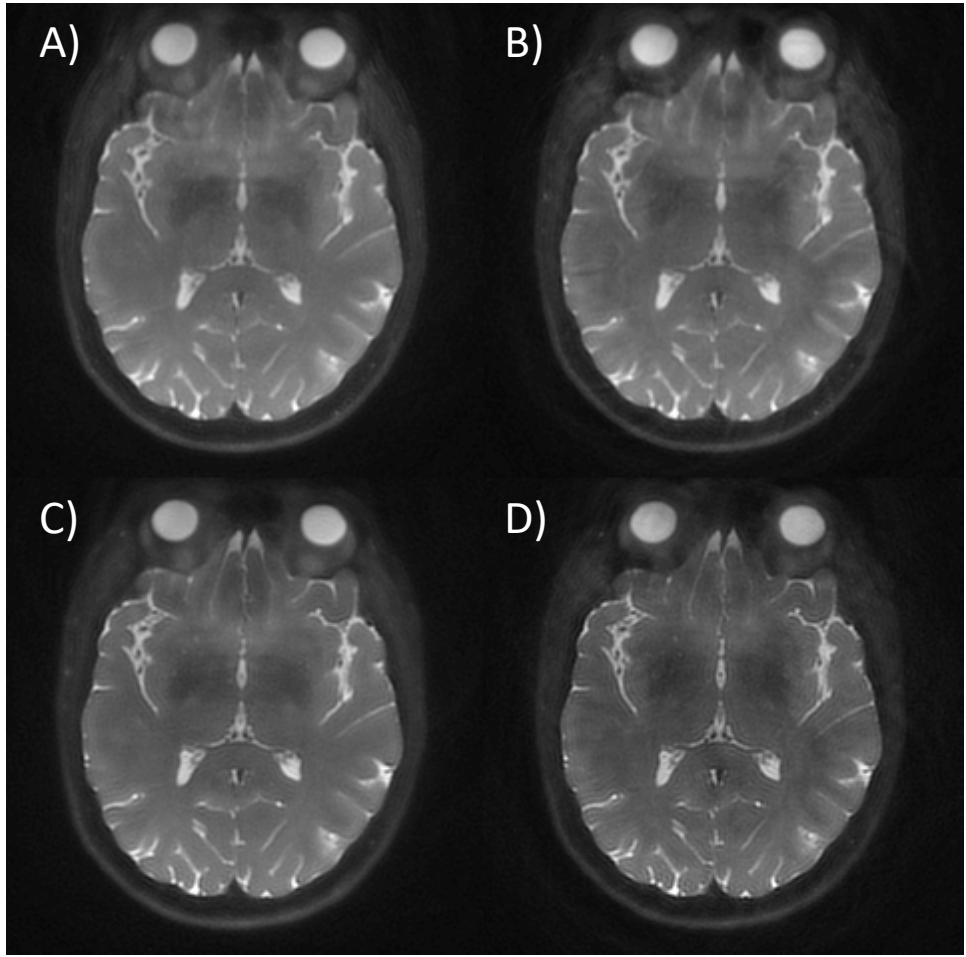

Figure 7. Single shot spiral TSE images acquired with fixed (A, B) and tangential (C, D) mode with continuous (A, C) and alternating (C, D) readout of the spiral segments.

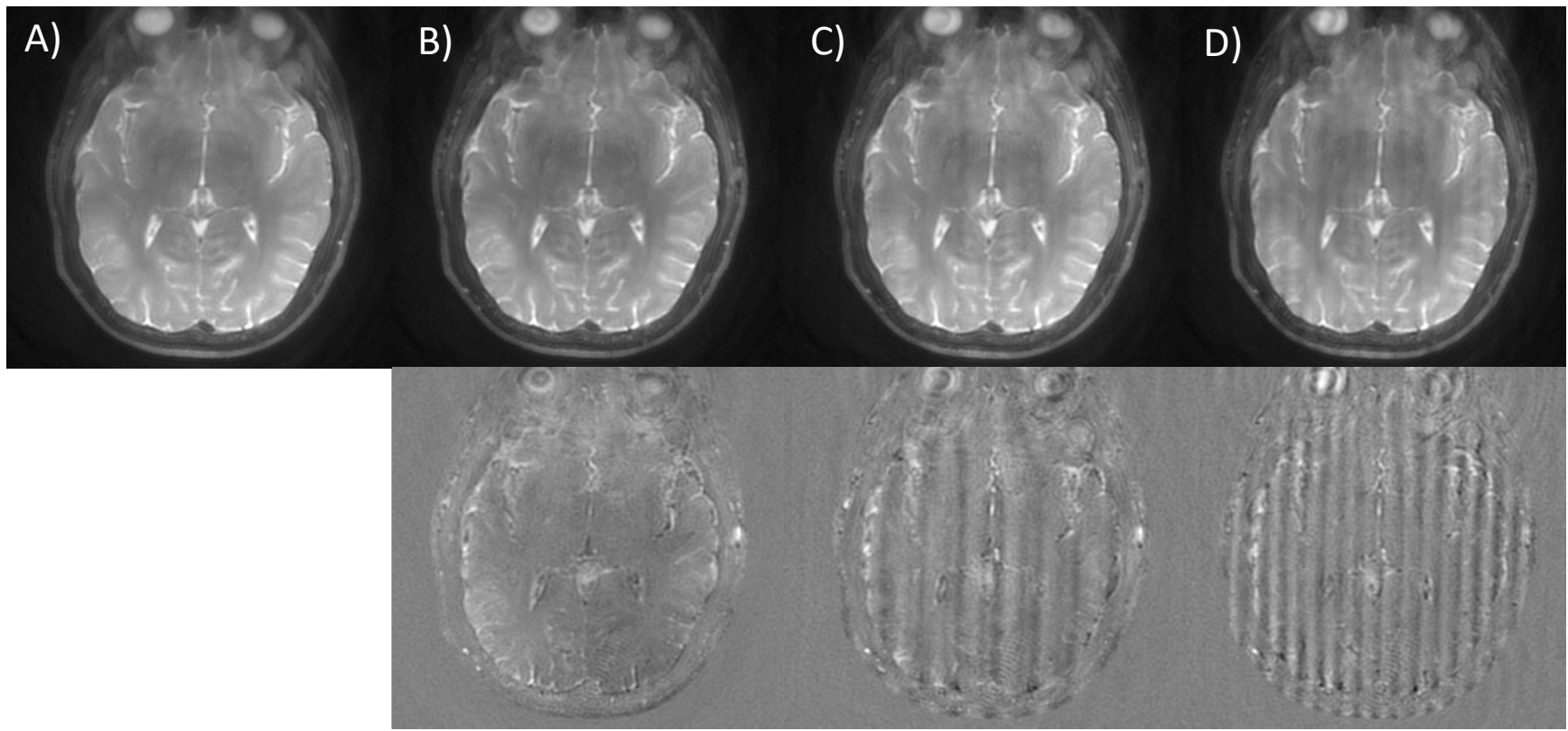

Figure 8. Spurious echoes interfere with the proper signals if $K_{ref+,-}$ is placed inside the k-space ranging from -500 to 500 1/m covered by the spiral trajectory for data acquisition with 1 mm spatial resolution; A) shows results of a scan with $K_{ref+,-}$ set to kx, ky= -550,0 1/m, B) for 0,0 (B), -50,0 (C), and -100,0 (D) 1/m, respectively. The bottom row represents difference images to A) scaled by a factor of 5 for better visualization.

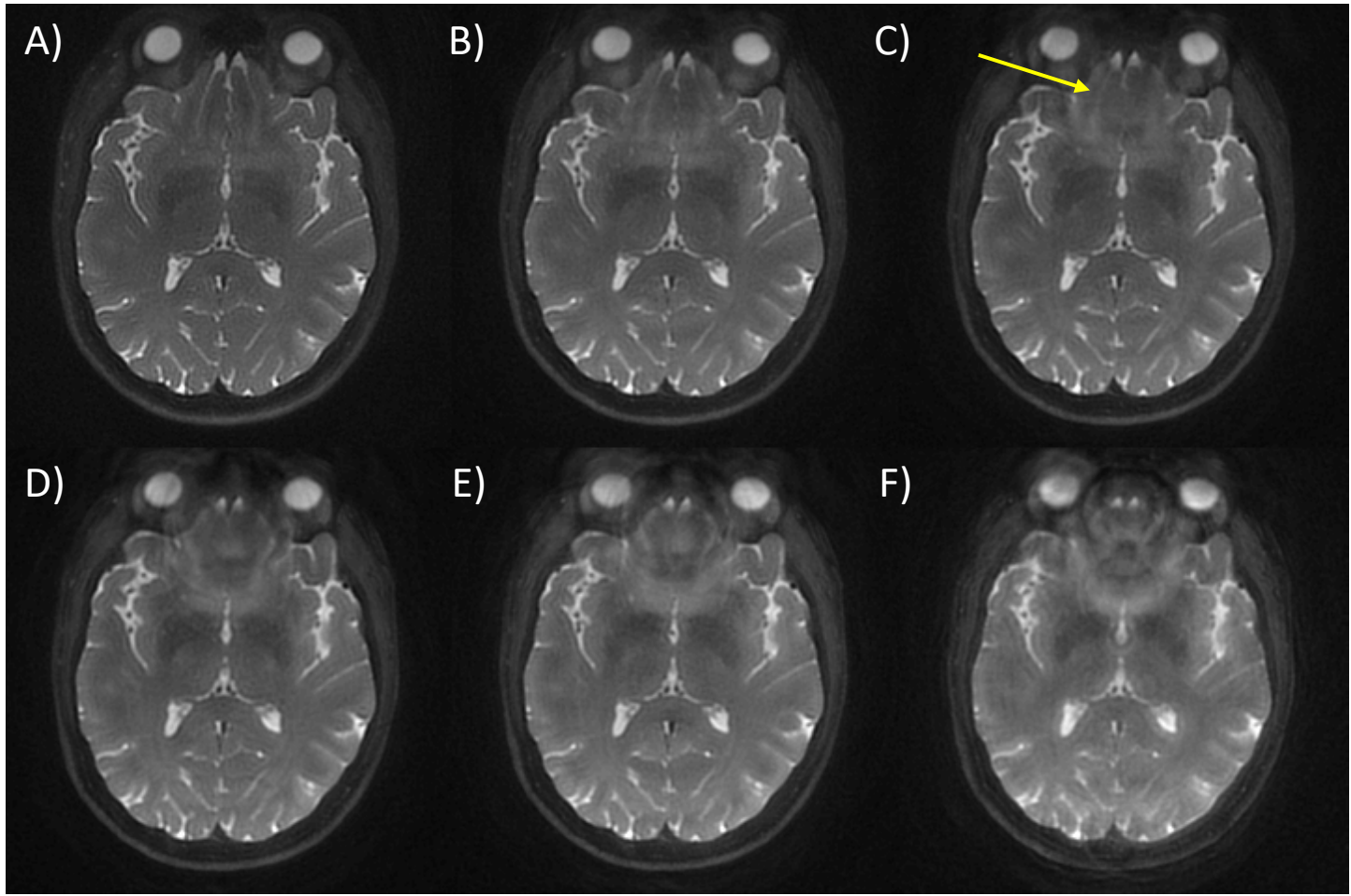

Figure 9. Images acquired with different echo spacing: A) 8 (3.9) ms, B) 10 (5.9) ms, C) 12 (7.9) ms, D) 14 (9.9) ms, E) 16 (11.9) and F) 20 (15.9) ms; numbers in brackets represent $t_{acq}$. Artifacts in the frontal brain above the sinuses indicated by the yellow arrow increase with increasing $t_{acq}$.

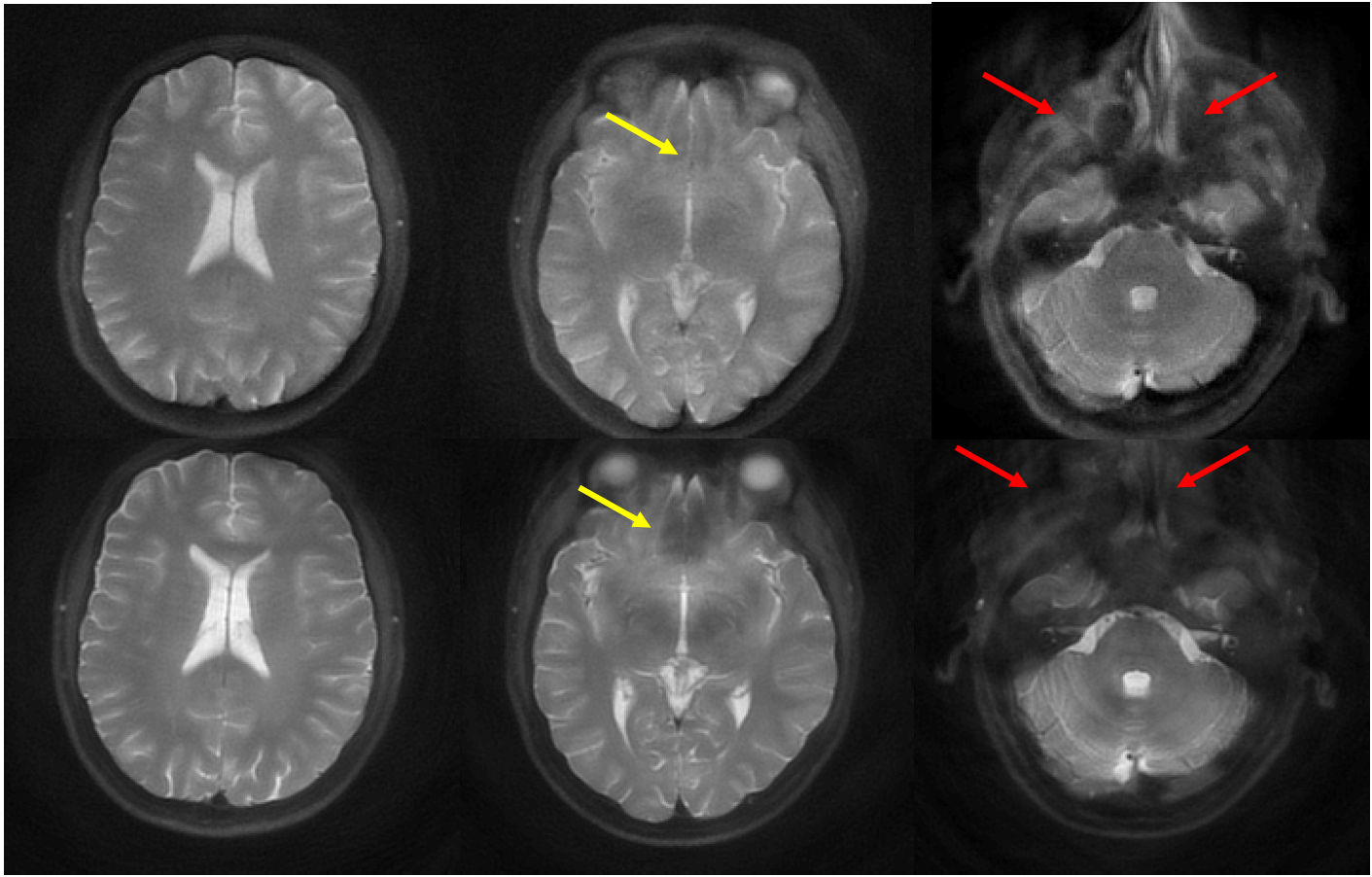

Figure 10. Comparison of spiral TSE images acquired at 1.5 T (top) and 3 T (bottom) on the same volunteer and with identical sequences. It is demonstrated that images at 1.5 T show reduced artifacts due to susceptibility effects in problematic regions (yellow arrows). Facial structures (red arrows) are better visible but are still strongly blurred at 1.5 T.

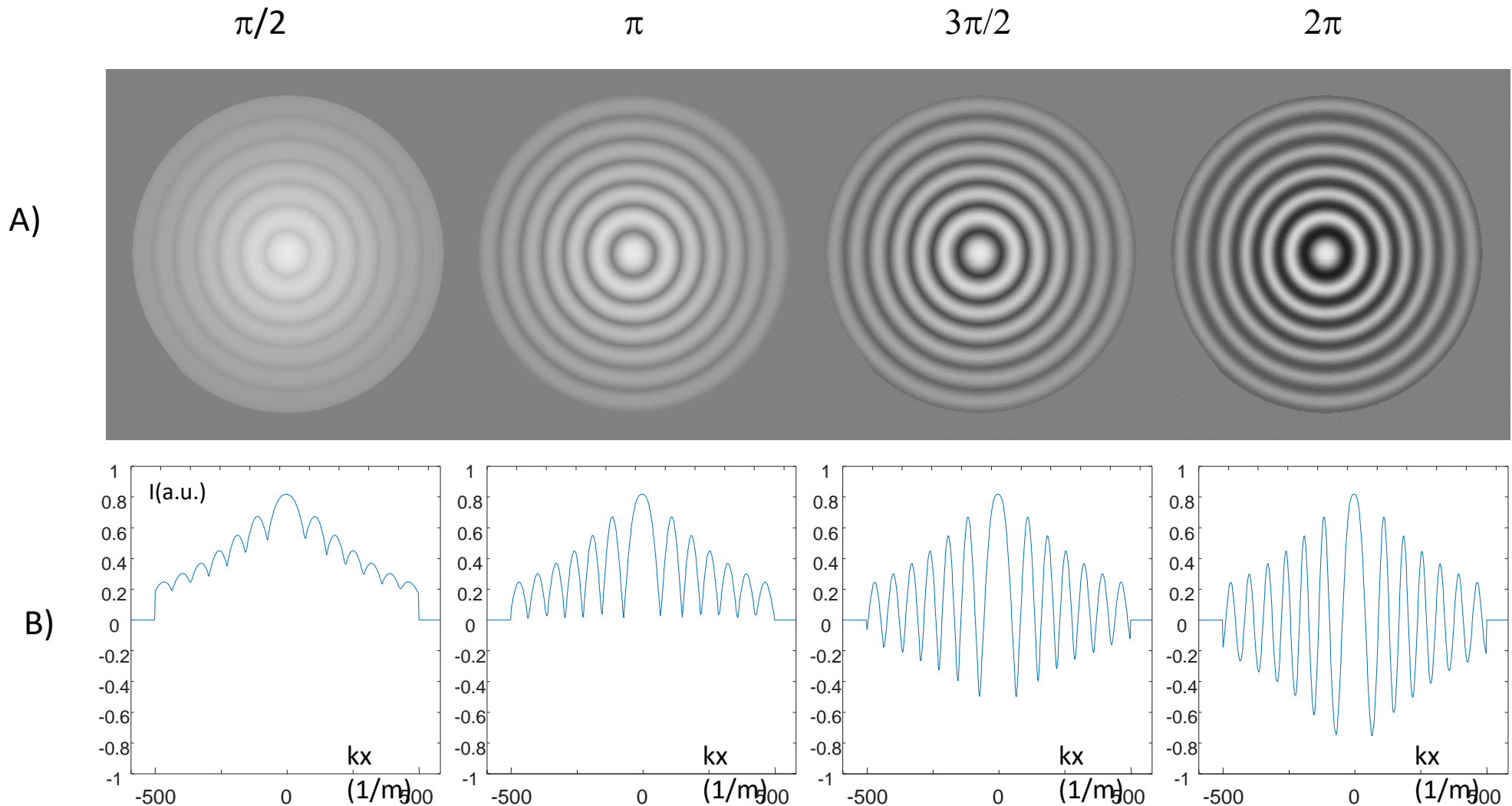

Figure S1. A) Modulation of k-space for annulated segmentation with 13 echoes as a function of the dephasing over $t_{acq}$. The real part of the modulation function is shown. The trajectory has been regridded to an image for better visualization. The annulated segmentation leads to a ring-like modulation. B) shows the modulation along a horizontal profile through the k-space center.

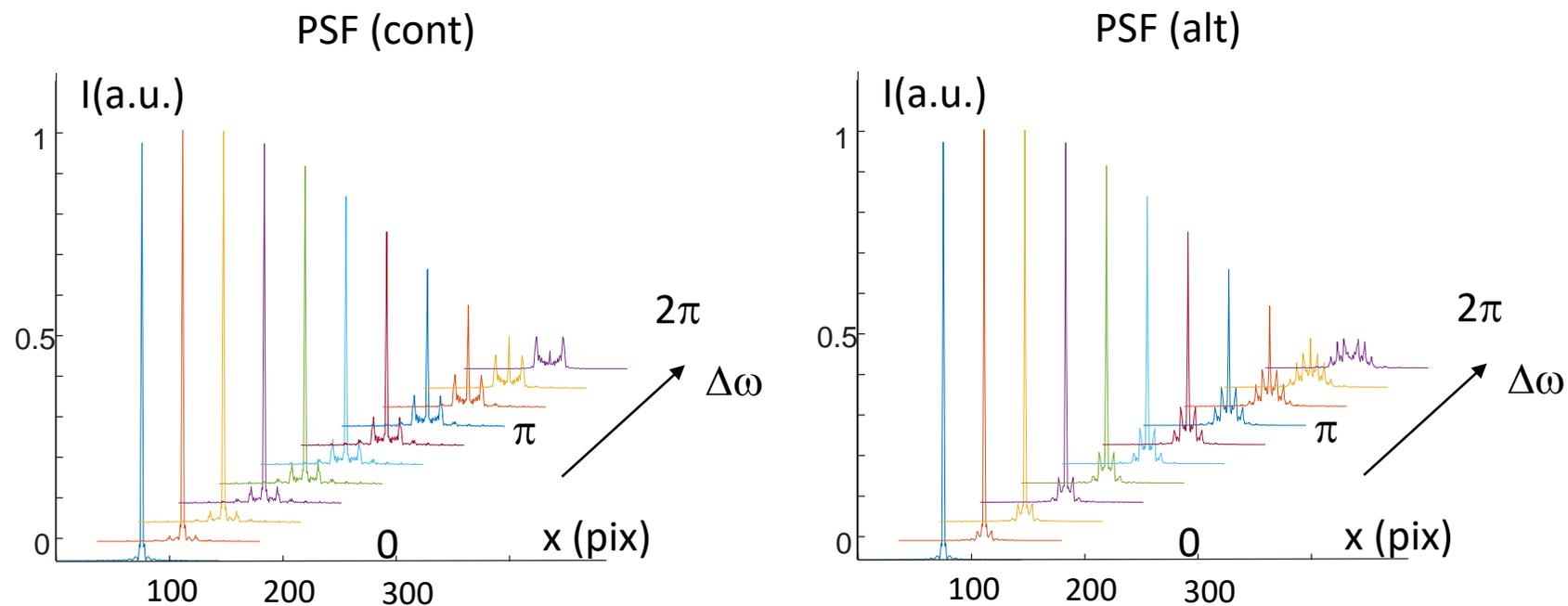

Figure S2. Comparison of the point spread function PSF for continuous (left) and alternating (right) acquisition mode. The signal intensity at the center is identical in both modes.

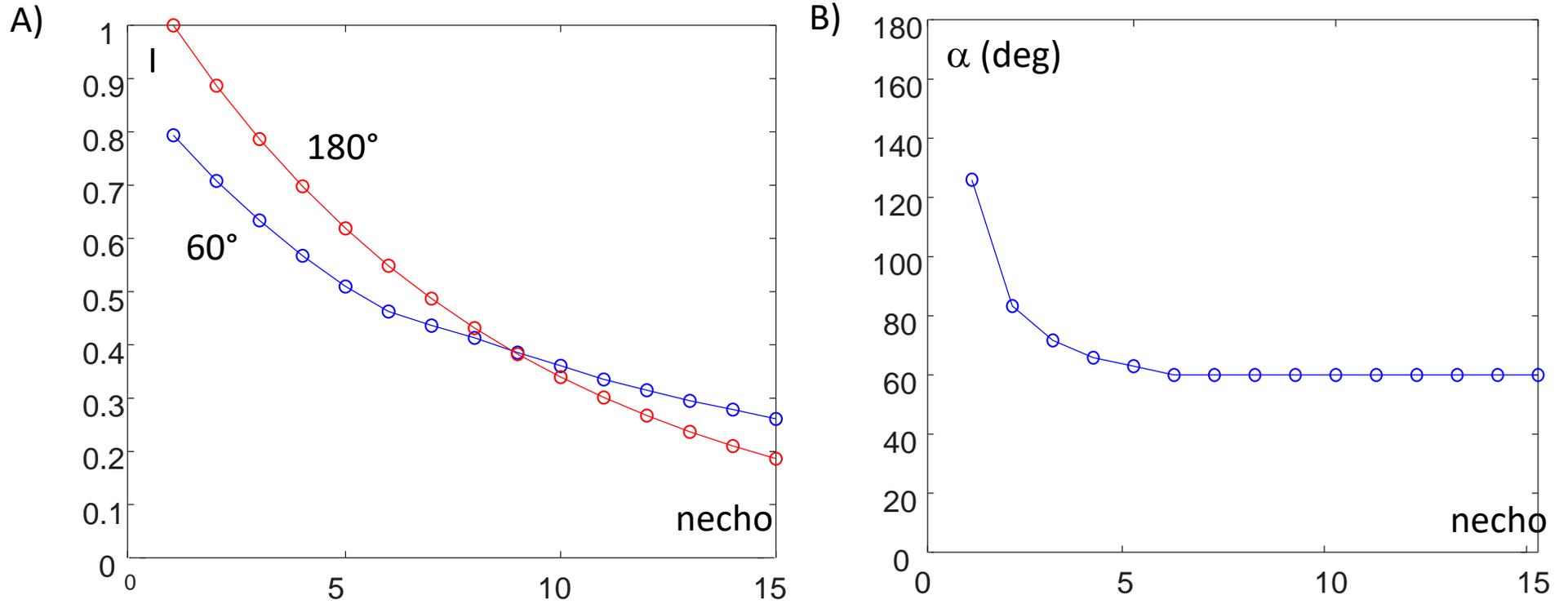

Figure S3. A) Signal decay over 15 echoes with ESP=10 ms for tissue with can be normalized with calculated amplitudes for image reconstruction for tissue with $T_1$=1 and $T_2$= 0.1 s. (B) Refocussing flip angles used for smooth transition into the pseudo steady state according to TRAPS. Specific absorption rate SAR is reduced to ~13%).

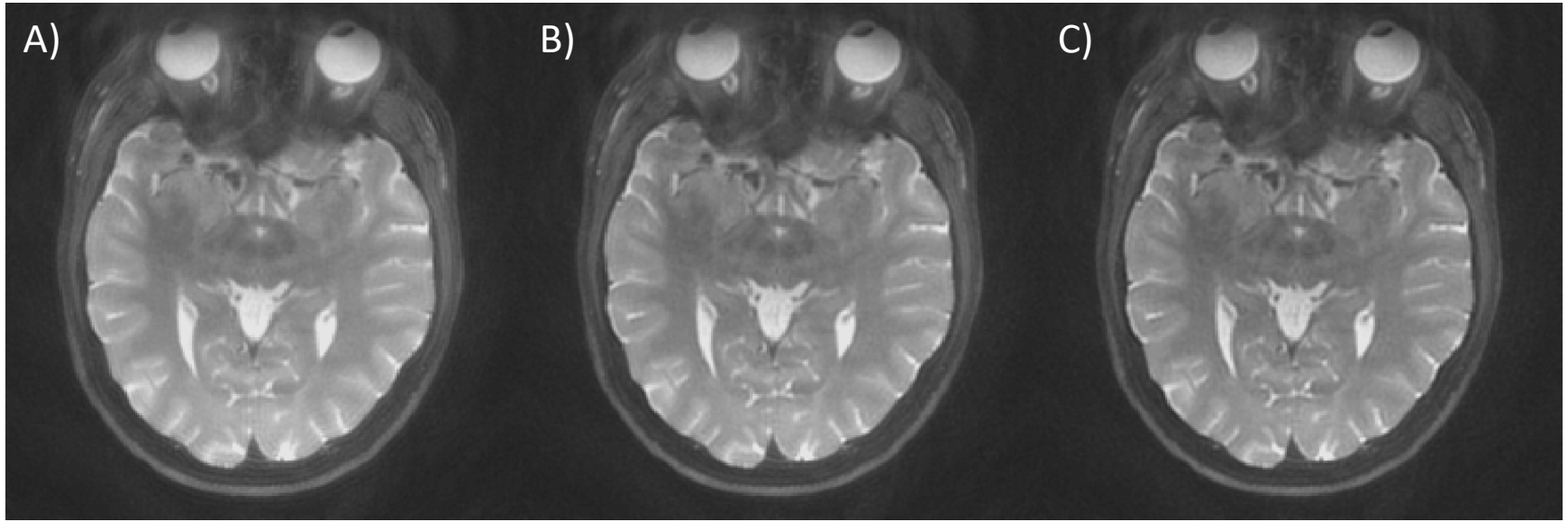

Figure S4. Improvement of image quality by the EPG-based normalization. The unfiltered image (A) acquired with refocusing flipangles 126°, 83°, 72°, 66°, 63°, 60°, 60°,….looks slightly blurred due to the initial signal decay compared to images reconstructed from data normalized with a simulated signal decay with T1= 1 s and T2= 200 ms (B) and T1=1s and T2=100 ms (C). The latter roughly corresponds to relaxation parameters for white matter at 3 T (37). The median values (B) were typically used in order to avoid artificial sharpening and/or Gibbs ringing of CSF-filled structures.

|  | Tangential mode | | | Fixed mode | | |
|---|---|---|---|---|---|---|
| ESP (ms) | $t_{acq}$ (ms) | ETL | ETD (ms) | $t_{acq}$ (ms) | ETL | ETD (ms) |
| 9 | 4.9 | 37 | 333 | 5.8 | 25 | 225 |
| 9.5 | 5.4 | 32 | 304.5 | 6.3 | 22 | 209 |
| 10 | 5.9 | 27 | 270 | 6.8 | 20 | 200 |
| 11 | 6.9 | 22 | 242 | 7.8 | 17 | 187 |
| 12 | 7.9 | 18 | 216 | 8.8 | 15 | 180 |
| 13 | 8.9 | 16 | 208 | 9.8 | 13 | 169 |
| 14 | 9.9 | 14 | 196 | 10.8 | 12 | 168 |
| 15 | 10.9 | 12 | 180 | 11.8 | 11 | 165 |
| 16 | 11.9 | 11 | 176 | 12.8 | 10 | 160 |
| 17 | 12.9 | 10 | 170 | 13.8 | 9 | 153 |
| 18 | 13.9 | 9 | 162 | 14.8 | 8 | 144 |
| 19 | 14.9 | 9 | 171 | 15.8 | 8 | 152 |
| 20 | 15.9 | 8 | 160 | 16.8 | 7 | 140 |

Supplementary Table 1: Acquisition parameters $t_{acq}$, echo train length ETL and echo train duration ETD (= ETL * $t_{acq}$) for acquisition of a variable density spiral with spoiler times $t_{sp}$ set to the maximum possible value (1.05 ms for fixed mode and 0.6 ms for tangential mode).